# A Software-Based Approach for Acoustical Modeling of Construction Job Sites with Multiple Operational Machines

Behnam SHERAFAT[1], Abbas RASHIDI[2], and Siyuan SONG[3]

[1] Ph.D. Student of construction engineering, Department of Civil and Environmental Engineering, University of Utah, Salt Lake City, Utah 84112; PH (801) 558-1001; FAX (801) 585-5477; email: behnam.sherafat@utah.edu

[2] Assistant Professor, Department of Civil and Environmental Engineering, University of Utah, Salt Lake City, Utah 84112; PH (801) 581-3155; FAX (801) 585-5477; email: abbas.rashidi@utah.edu

[3] Assistant Professor, School of Construction and Design, The University of Southern Mississippi, Hattiesburg, MS 39406; PH (601) 266-6259; FAX (601) 266-5717; email: siyuan.song@usm.edu

**ABSTRACT**

Several studies have been conducted to automatically recognize activities of construction equipment using their generated sound patterns. Most of these studies are focused on single-machine scenarios under controlled environments. However, real construction job sites are more complex and often consist of several types of equipment with different orientations, directions, and locations working simultaneously. The current state-of-research for recognizing activities of multiple machines on a job site is hardware-oriented, on the basis of using microphone arrays (i.e., several single microphones installed on a board under specific geometric layout) and beamforming principles for classifying sound directions for each machine. While effective, the common hardware-approach has limitations and using microphone arrays is not always a feasible option at ordinary job sites. In this paper, the authors proposed a software-oriented approach using Deep Neural Networks (DNNs) and Time-Frequency Masks (TFMs) to address this issue. The proposed method requires using single microphones, as the sound sources could be differentiated by training a DNN. The presented approach has been tested and validated under simulated job site conditions where two machines operated simultaneously. Results show that the average accuracy for soft TFM is 38% higher than binary TFM.

**INTRODUCTION**

Construction automation has been widely used in different construction areas such as detecting concrete rebar (Xiang and Rashidi 2019, Xiang et al. 2019), building quantity take-off (Shafaghat et al. 2019, Taghaddos et al. 2016), and construction site path planning (Song and Marks 2019). Also, automating construction performance monitoring is a continual and critical conversation to be had in the construction domain (Asadi et al. 2019a, Asadi et al. 2019b) and automatically recognizing activities of construction equipment using the generated sound patterns is one of the recent subjects in this domain. Several studies have been conducted on this subject, which use different types of microphones such as single microphones and microphone arrays. Most of these studies are focused on detecting and recognizing activities of single-machine scenarios (Cheng et al. 2017a, Cheng et al. 2017b, Sabillon et al. 2017, Sherafat et al. 2019, Zhang et al. 2018). However, real construction job sites consist of multiple machines working simultaneously, which is more challenging than single-machine scenarios (Cheng et al. 2019). Recognizing activities of these machines requires separating different sound sources using Blind Source Separation (BSS) algorithms and applying single-machine activity recognition methods.

BSS is a partly developed research topic in several application areas (Cardoso 1998), aimed at recovering source signals from a set of mixed signals, with very little information about the source signals or the mixing process. BSS can be classified as follows (Comon 2004, Niknazar et al. 2014): 1) under-determined BSS (when number of sensor signals are less than that of sources); 2) determined (when number of sensor signals equal number of sources); and 3) over-determined (when number of sensor signals are more than that of sources).

The classical application of blind source separation is the cocktail party problem, referring to several people are talking simultaneously and a listener trying to follow a single monologue (Qian et al. 2018). Different methods are available for BSS and some of them are as follows: 1) Principal Components Analysis (PCA) (Serviere and Fabry 2005); 2) Singular Value Decomposition (SVD) (De Lathauwer et al. 1994); 3) Independent Component Analysis (ICA) (Saruwatari et al. 2003); 4) Non-negative Matrix Factorization (NMF) (Cichocki et al. 2006, Févotte et al. 2018). Moreover, some other applications of BSS are as follows: 1) male/female speech separation; 2) speech/music separation; 3) musical notes separation; and 4) vocal/non-vocal music separation.

There are two types of approaches: 1) hardware-approach: This method needs special hardware settings such as microphone arrays and also a wired connection to a laptop to save the data. Microphones placed on this device record different mixed signals with their related delays. Using over-determined BSS algorithms, different sound sources can be separated; 2) software-approach: This method only needs one single channel microphone placed on the job site without any other hardware requirements. This paper contributes to the body of knowledge by introducing a method based on deep learning, capable of separating equipment sound sources on the construction site. The results of this paper can later be used for construction equipment activity recognition using their generated sounds.

**LITRETURE REVIEW**

BSS has a broad background in other research areas (e.g., music) other than construction area. Several studies have been conducted on this topic and different types of methods and algorithms are introduced. In one of the most recent studies, Sun et al. (2019) proposed a new optimization function of joint-dictionary learning, based on identity sub-dictionaries and common sub-dictionary for single-channel blind separation source. Their method trains these two types of dictionaries, which considers both the uniqueness and similarity between speech signals of males and females. Smith et al. (2019) utilized an iterative algorithm, named "Iterative Least Squares with Enumeration (ILSE)", with different moment-based mixing vector estimators as initializers to fully exploit sources' geometry. DNNs have been utilized to estimate the ideal hard/soft masks to separate sound sources from a mixed signal. Moreover,

Simpson et al. (2015) used DNN to extract vocals from musical mixtures. DNN has been used to estimate ideal binary masks. Binary masks are matrixes of 0 and 1, where each element of this matrix is determined by comparing the magnitude values of sound sources from their corresponding spectrogram.

Recently, a few studies were focused on the application of sound source separation algorithms in the construction industry. Cheng et al. (2019) used Steered Response Power and delay-and-sum beamforming to separate different equipment sound sources while recognizing their activities. However, they have focused on the activity recognition aspect of the research rather than the sound source separation. Thus, in this paper, we have proposed a method using DNN to separate multiple operational machine sound sources on the construction site. This method resolves the issues of the hardware-oriented approach, which is based on using microphone arrays (requires connecting to a laptop) and uses the recorded sounds of a single microphone. The authors evaluated the performance of this method under artificial and realistic job site conditions where multiple machines operated simultaneously. Also, a comparison between this method and statistical and computational techniques (i.e., ICA and PCA) is provided.

**METHODOLOGY**

Separating equipment sound sources on the job site is partially similar to the Cocktail Party problem. Both the cocktail party and construction site have significant noise, which may interfere with the desired signals. Both the speech signal and the equipment signal have trackable patterns in the time and frequency domain, which can be utilized for training a machine learning model. The object of this paper is to separate individual machines from a mixed construction sound using a DNN, where two types of machines are working simultaneously.

One of the most effective methods for separating sound sources from a mixed signal is using the TFMs. A TFM is a matrix of the same size as the corresponding Short Time Fourier Transform (STFT) consisting of binary (0 and 1) values (binary/hard mask) or values in the range (0,1) (soft mask) (Wang 2008). This mask is multiplied element-by-element with its corresponding STFT to separate the sound sources. The idea behind calculating binary TFM is when the power of the excavator sound is greater than the power of the jackhammer sound at a specific time-frequency cell, the binary TFM's value will be set to 1 (otherwise being set to zero). On the other hand, the value of soft TFM is equal to the ratio of the excavator sound power to the total mixed signal power.

**Mask Estimation Using DNN**

Each cell in TFM represents if the cell is excavator dominant or jackhammer dominant. In other words, it proves that this cell contains more sound energy from an excavator or jackhammer. Based on this binary cell (existence of each sound) or soft cell (the probability of existence of each sound), it separates the two different sounds. The TFMs and DNN for sound source separation were used together for estimation purposes. The framework of DNN training is shown in Figure 1. The predictor is the magnitude spectra of the mixed signals and the target is the TFM corresponding to the excavator. J and E show the magnitude spectra of jackhammer and excavator, respectively. DNN uses the predictor to minimize the mean square error between the input and output values of the target. To generate the separated excavator sound, the output magnitude spectrum and the phase of the mixed signal are used together to convert the time-frequency representation of the signal to the time domain using inverse STFT. Details of the framework are elaborated in the following sections.

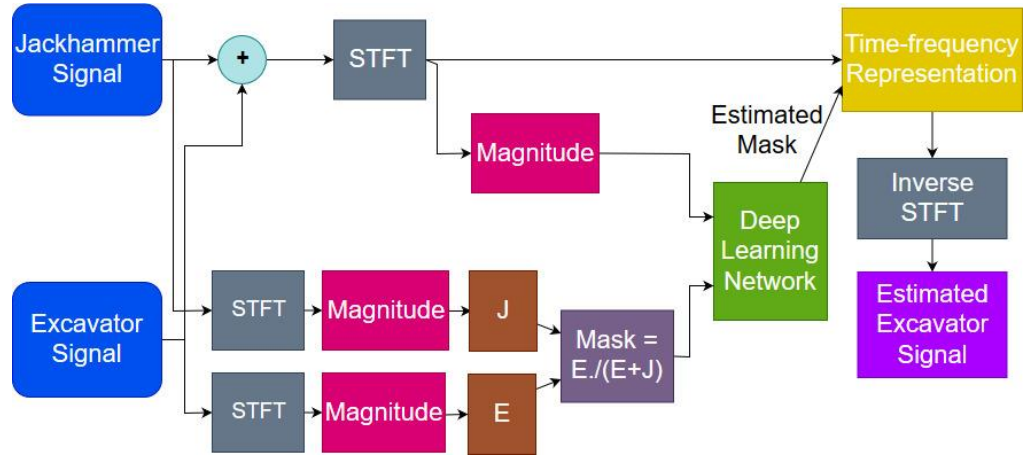

**Figure 1. The framework of excavator sound separation using soft TFM.**

*Read Training and Validation Signals*

First, the 60-seconds duration of machine sound is loaded into MATLAB. MATLAB is a powerful software for DNN and it has all of the required packages and toolboxes. 90% of this duration (i.e., 54 seconds) is used for training and 10% of that (i.e., 6 seconds) is used for validation. These signals are recorded using a single microphone and their sampling frequency is 44.1 KHz. In this paper, the authors have artificially combined the sound sources to compare the estimated sound sources with the actual sound sources, using performance measurement indexes. These signals are scaled to have the same power. From there. The sound sources are mixed to create the training and validation signals. In Figure 2, 40 seconds of the original (excavator and jackhammer) and mixed signals are visualized.

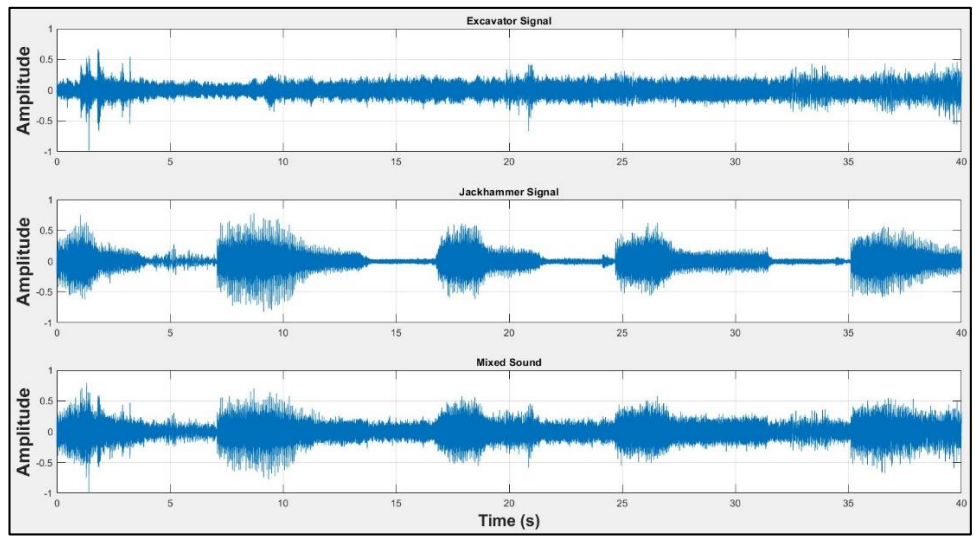

**Figure 2. Original (excavator and jackhammer) signals and mixed signals.**

*Generate Training and Validation STFTs*

Time-frequency representation of both signals and the mixed signal (both training and validation sets) are calculated using a Hanning window with a length of 128, an overlap length of 127, and an FFT length of 128. In Figure 3, time-frequency representations of excavator, jackhammer, and mixed signals are shown. Finally, the logarithm of the STFTs is taken and they are normalized.

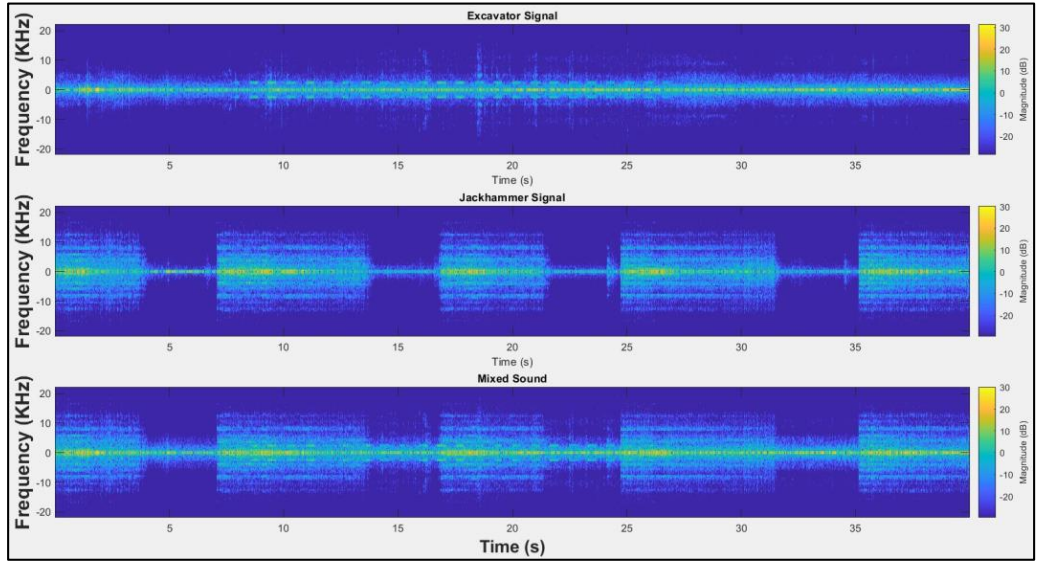

**Figure 3. Time-frequency representation of original (excavator and jackhammer) signals and mixed signals.**

*Compute the Training and Validation Soft Masks*

Matrix of training and validation soft masks are calculated using the Equation (1).

$$TFM = \frac{Exavator\ STFT\ Magnitude}{(Exavator\ STFT\ Magnitude + Jackhammer\ STFT\ Magnitude)} \quad (1)$$

*Deep Learning Network*

Chunks of size (65, 20) with an overlap length of 10 for both predictor and target signals are created. The layers of the network are defined as follows: 1) input layer of size 1×1× (65×20); 2) two hidden fully connected layers with 1300 neurons followed by sigmoid activation function; and 3) an output fully connected layer with 1300 neurons, followed by a regression layer. The number of epochs and the batch size (training signals at a time) is defined as 3 and 64, respectively. The training sample is shuffled at each epoch and the network uses Adaptive Moment Estimation (ADAM) solver. The estimated soft mask is multiplied element-by-element, with its corresponding STFT to separate the excavator sounds. Finally, the result of this multiplication (i.e., excavator time-frequency representation) is converted to time-domain using inverse STFT to show the excavator time domain signal.

**RESULTS**

The authors chose to use the jackhammer and excavator in this approach, due to the fact that these two machines have different patterns. For example, jackhammer has a repeating pattern and excavator has a uniform pattern. Thus, it helps create a better visual understanding when the original and estimated signals are compared. The sounds of both machines are recorded using a single channel microphone in an actual job site. The estimated jackhammer and excavator signals for binary TFM and soft TFM are shown in Figure 4 and 5. Figure 4 shows the original and estimated signals for both machines using the binary TFM method. Figure 5 shows the original and estimated signals for both machines using the soft TFM method. The original and estimated signals have similar outlines, which shows that the proposed method is efficient in separating two signals. Also, to measure the performance of source separation, authors have used BSS Eval toolbox in MATLAB (Vincent et al. 2006, Vincent et al. 2012). This toolbox decomposes the estimated signal to several components (Equation 2).

$$\widehat{S_j} = S_{target} + e_{interf} + e_{noise} + e_{artif} \qquad (2)$$

Where $S_{target}$ is a distorted version of $s_j$ (original signal) by an allowed distortion, and where $e_{interf}$, $e_{noise}$, and $e_{artif}$ are the interferences, noise, and artifacts error terms, respectively. Based on estimated components, several numerical performance criteria are calculated by computing energy ratios expressed in decibels (dB) (Equation 3, 4, 5).

Source to Distortion Ratio (SDR):

$$SDR = 10 \log_{10} \frac{\|S_{target}\|^2}{\|e_{interf} + e_{noise} + e_{artif}\|^2} \qquad (3)$$

Source to Interferences Ratio (SIR):

$$SIR = 10 \log_{10} \frac{\|S_{target}\|^2}{\|e_{interf}\|^2} \qquad (4)$$

Source to Artifacts Ratio (SAR):

$$SAR = 10 \log_{10} \frac{\|S_{target} + e_{interf} + e_{noise}\|^2}{\|e_{artif}\|^2} \qquad (5)$$

These values are calculated for TFM (both binary and soft) and ICA, and the results are shown in Table 1. FastICA is proved to be an efficient and popular algorithm for independent component analysis (Koldovsky et al. 2006). This algorithm maximizes a measure of non-gaussianity of the rotated components and uses kurtosis and negentropy, which are the measures of the tailedness and distance to normality, respectively.

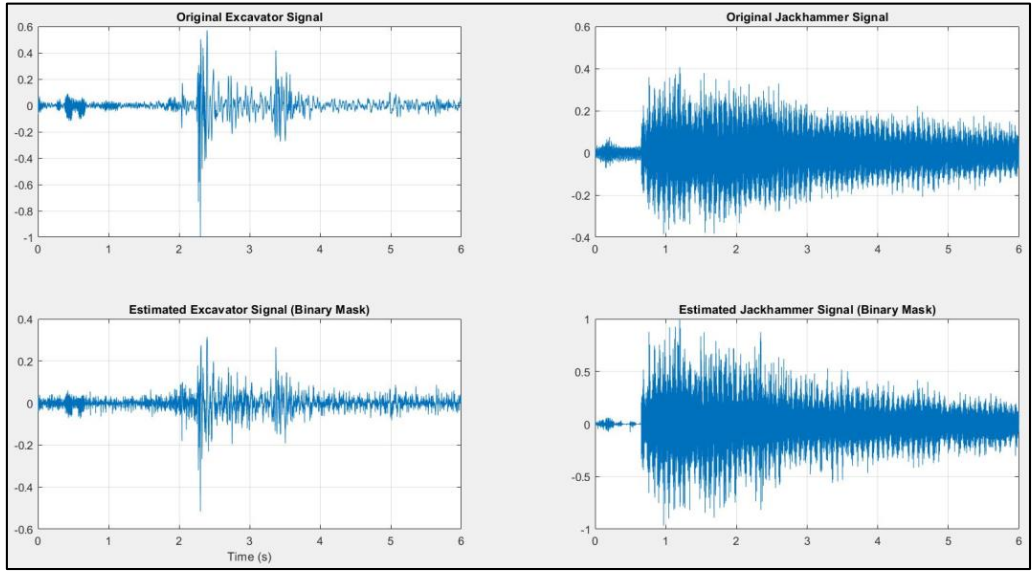

**Figure 4. The output of sound separation for the binary TFM.**

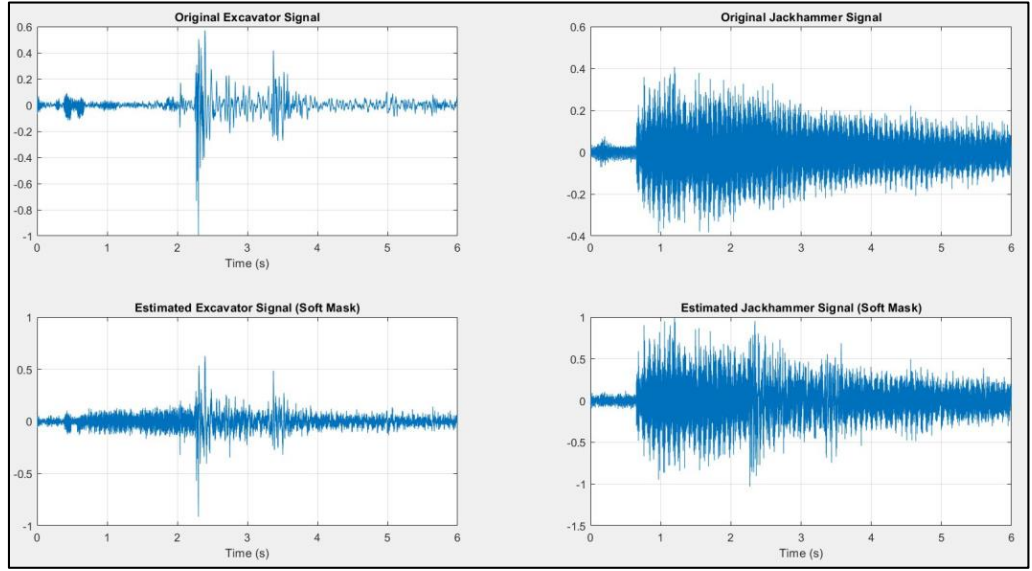

**Figure 5. The output of sound separation for the soft TFM.**

**Table 1. Evaluation of source separation using binary and soft TFM.**

| Method | SDR (dB) | | SIR (dB) | | SAR (dB) | |
|---|---|---|---|---|---|---|
| | $\hat{E}$ | $\hat{J}$ | $\hat{E}$ | $\hat{J}$ | $\hat{E}$ | $\hat{J}$ |
| Binary TFM | -9.30 | 10.90 | -8.10 | 18.30 | 5.50 | 11.80 |
| Soft TFM | -8.90 | 15.10 | -8.70 | 17.80 | 15.20 | 18.50 |
| FastICA Negentropy | 0.03 | 0.04 | 0.03 | 0.04 | 61.15 | 61.15 |
| FastICA Kurtosis | 0.03 | 0.04 | 0.03 | 0.04 | 61.15 | 61.15 |

## CONCLUSION

Comparing the results for different methods shows that TFM has promising results with respect to the FastICA algorithm. FastICA is a numerical algorithm and uses a computational technique to estimate independent components. This algorithm lacks signal processing methods and does not consider the frequency content of signals. For this reason, SDR, SIR, and SAR are the same for both negentropy and kurtosis. Moreover, soft TFM has 18.1%, 0.4%, 94.8% higher performance than binary TFM for SDR, SIR, and SAR, respectively.

The output of this paper can be used for construction equipment activity recognition. A single microphone can record sounds of multiple machines on the jobsite. Then, different sound sources are separated and activity recognition algorithms can be applied on each of the sound signals. This study is a one of the initial steps toward construction equipment performance monitoring using generated sounds.

In future studies, authors will apply this method to more realistic scenarios. For example, this method requires generalization for when three or more machines are working simultaneously. Also, performance measurement parameters used in this paper require the original sound sources. Hence, in this paper, an artificial mix of the signals has been used. In future studies, realistic mixing and the corresponding performance measurements need to be investigated. Currently, this method is tested on different types of machines. In the future, authors will evaluate this method for scenarios when different machines of the same type (e.g., different excavators) are working simultaneously.

## ACKNOWLEDGMENTS

This research project has been funded by the U.S. National Science Foundation (NSF) under Grant CMMI-1606034. The authors gratefully acknowledge NSF's support. Any opinions, findings, conclusions, and recommendations expressed in this manuscript are those of the authors and do not reflect the views of the funding agency.